\documentclass[a4,12pt]{article}
\usepackage{type1cm,amsmath,amssymb,color,graphics,amscd,amsfonts,mathrsfs,epsf,indentfirst}
\usepackage{epsfig}
\usepackage{bm}
\usepackage{subfigure}
\usepackage{multirow,graphicx}
\allowdisplaybreaks[1]

\makeatletter
\@addtoreset{equation}{section}

\makeatletter
\renewcommand\section{\@startsection {section}{1}{\z@}%
                                   {-3.5ex \@plus -1ex \@minus -.2ex}
                                   {2.3ex \@plus.2ex}%
                                   {\normalfont\large\bfseries}}
\renewcommand\subsection{\@startsection{subsection}{2}{\z@}%
                                     {-3.25ex\@plus -1ex \@minus -.2ex}%
                                     {1.5ex \@plus .2ex}%
                                    {\normalfont\bfseries}}

\begin{document}

\begin{titlepage}
  \thispagestyle{empty}

  \begin{flushright} 
    DAMTP-2011-21
  \end{flushright} 
    
  \vspace{2cm}
  
  \begin{center}
    \font\titlerm=cmr10 scaled\magstep4
    \font\titlei=cmmi10 scaled\magstep4
    \font\titleis=cmmi7 scaled\magstep4
     \centerline{\titlerm
Conformal weights in the
Kerr/CFT correspondence}
    
    \vspace{2.2cm}
    \noindent{
     Keiju Murata
      }\\
    \vspace{0.8cm}
    
   {\it DAMTP, University of Cambridge, Centre for Mathematical Sciences, \\ Wilberforce Road, Cambridge CB3 0WA, UK}
    
K.Murata@damtp.cam.ac.uk

   \vspace{1cm}
   {\large \today}
  \end{center}

  \vskip 3em

\begin{abstract}
It has been conjectured that a near-extreme Kerr black hole is described
 by a 2d CFT. Previous work has shown that CFT operators dual to
 axisymmetric gravitational perturbations have integer conformal
 weights. In this paper, we study the analogous problem in 5d. We
 consider the most general near-extreme vacuum black hole with two
 rotational symmetries. This includes Myers-Perry black holes, black
 rings and Kaluza-Klein black holes. We find that operators dual to
 gravitational (or electromagnetic or massless scalar field)
 perturbations preserving both rotational symmetries have integer
 conformal weights, the same for all black holes considered.
\end{abstract}

\end{titlepage}


\section{Introduction and summary}

The Kerr-CFT correspondence is a conjectured microscopic description for a Kerr black hole~\cite{Guica:2008mu,Bredberg:2011hp}. Originally it was proposed that quantum gravity in the near-horizon extreme Kerr (NHEK \cite{Bardeen:1999px}) geometry is described by a chiral 2d CFT. Surprisingly, it is the $U(1)$ of the $U(1)_L \times SL(2,R)_R$ isometry group that gets enhanced to a Virasoro symmetry in this proposal. States that transform non-trivially under $SL(2,R)_R$ are expected to correspond to excitations away from extremality. Therefore it is natural to ask whether there exists a non-chiral CFT, in which $SL(2,R)_R$ also is enhanced to a Virasoro symmetry, that describes near-extreme Kerr black holes.

The decoupling limit of a near-extreme Kerr gives a "near-NHEK" geometry which is locally isometric to NHEK \cite{Dias:2009ex,Amsel:2009ev}. One might try to define boundary conditions in the (near-)NHEK geometry for which the asymptotic symmetry group contains two sets of Virasoro generators. Quantum gravity with such boundary conditions would be a non-chiral CFT.\footnote{For work in this direction, see Refs.~\cite{Matsuo:2009sj,Matsuo:2009pg,Castro:2009jf,Rasmussen:2010sa}.} Unfortunately, Refs. \cite{Dias:2009ex,Amsel:2009ev} argued that the gravitational backreaction of any perturbation would destroy any such boundary conditions.\footnote{
The results described in this paragraph were obtained earlier for Reissner-Nordstrom black holes \cite{Maldacena:1998uz}.}

Nevertheless, evidence that a non-chiral CFT description for near-extreme Kerr does exist is provided by studies of scattering \cite{Bredberg:2009pv,Cvetic:2009jn,Hartman:2009nz}. A CFT dual to the NHEK geometry should describe modes close to the superradiant limit: $M|\omega - m\Omega| \ll 1$ (where $M$ is the black hole mass and $\Omega$ its angular velocity, and the waves have frequency $\omega$ and azimuthal quantum number $m$). It is found that the scattering cross-section for such waves from near-extreme Kerr can be reproduced from correlation functions in a 2d non-chiral CFT.\footnote{In a separate development, it has been observed that the scattering cross-section for low-frequency ($M \omega \ll 1$) waves from a general Kerr black hole also can be reproduced from 2d non-chiral CFT correlation functions \cite{Castro:2010fd}.}

2d CFT operators are classified by their conformal weights $(h_L,h_R)$. $h_R$ is associated to $SL(2,R)_R$, which is realized geometrically in NHEK. Therefore one can use AdS/CFT techniques to calculate $h_R$ for an operator dual to a given mode of a bulk field. One performs a Kaluza-Klein reduction on the $S^2$ or NHEK to reduce a given bulk field to a tower of fields in $AdS_2$. From the $SL(2,R)$ transformation properties of these fields one can read off the weights $h_R$ for dual CFT operators. Given $h_R$, Ref. \cite{Hartman:2009nz} found that taking $h_L = h_R - |s|$ reproduces the scattering cross-section for a field of helicity $s$.

In carrying out this procedure, one must ask which fields to include in the bulk. Since we don't know the correct theory of quantum gravity in the bulk, the most conservative approach is to focus on the only field that any such theory must include, namely the gravitational field. So we shall consider operators dual to linearized gravitational perturbations. These were studied in Refs. \cite{Dias:2009ex,Amsel:2009ev}. KK reduction on $S^2$ leads to an infinite set of fields in $AdS_2$ labelled by integers $(l,m)$ ($|m| \le l=2,3,\ldots$) corresponding to spheroidal harmonics on $S^2$. For each field one can read off $h_R$. 
For modes with $m \ne 0$, the result is complicated, involving an eigenvalue that must be determined numerically. Modes with $|m| \sim l$ give complex $h_R$ \cite{Bardeen:1999px,Dias:2009ex,Amsel:2009ev}. The microscopic interpretation of this is unclear. 

However, for axisymmetric modes, i.e., those with $m=0$, one finds that $h_R$ is an integer: $h_R = l+1$. Hence there exists an infinite family of CFT operators invariant under $U(1)_L$ with integer conformal weights. The same result arises for CFT operators dual to KK harmonics of a massless scalar field or an electromagnetic field.\footnote{The proposal of Ref. \cite{Castro:2010fd} also gives integer conformal weights for such fields, even for $m \ne 0$.} We are not aware of any explanation of why these conformal weights should be integers.\footnote{Note that the operator of lowest weight dual to these bulk fields has $h_R=3$ whereas any CFT should contain an operator with $h_R=2$, namely the right-moving part of the energy momentum tensor $\tilde{T}$. One might expect $\tilde{T}$ to be dual to bulk modes that correspond to a change in mass of the mass of the black hole. However, such modes are non-dynamical  \cite{Dias:2009ex,Amsel:2009ev}, i.e., they do not correspond to bulk fields.} 

The purpose of this paper is to observe that this unexplained fact appears to be quite general. We shall consider 5d near-extreme vacuum black holes. As in 4d, the near-horizon geometry of such a black hole is locally isometric to the near-horizon geometry of the corresponding extreme black hole. All known examples of 5d black holes have two rotational symmetries. Any 5d near-horizon geometry with two rotational symmetries must take the form of a fibration over AdS$_2$ \cite{Kunduri:2007vf}
\begin{equation}
 ds^2=L(\theta)^2\left(-r^2 dt^2 +\frac{dr^2}{r^2}\right)
+\alpha(\theta)^2 d\theta^2
+\gamma_{IJ}(\theta)(d\phi^I-k^I r dt)(d\phi^J-k^J r dt)
\ ,
\label{NH4D5D}
\end{equation}
where $I,J=1,2$, $\partial_{\phi^I}$ are the rotational Killing vector fields of
the black hole and $k^I$ are constants. The metric has $U(1)\times U(1) \times SL(2,R)$ isometry group.
In fact, all vacuum near-horizon geometries of this form are known~\cite{Kunduri:2008rs,Hollands:2009ng}. They fall into three classes: (i) the 3-parameter near-horizon geometry of an extreme boosted Kerr black string (wrapped on a KK circle), which includes the near-horizon geometry of an extreme black ring \cite{Pomeransky:2006bd} as a special case; (ii) 2-parameter near-horizon extreme Myers-Perry \cite{Myers:1986un}, which is the same as the near-horizon geometry of a "slow" extreme Kaluza-Klein black hole \cite{Rasheed:1995zv,Larsen:1999pp}; (iii) the 3-parameter near-horizon geometry of a "fast" extreme KK black hole.

To compute operator dimensions in these geometries, we shall make use of recent results concerning decoupling of gravitational perturbations in near-horizon geometries. Ref. \cite{Durkee:2010qu} obtained a decoupled equation for a local gauge-invariant quantity describing such perturbations. Ref.  \cite{Durkee:2010ea} showed that one can KK reduce this equation on the space parameterized by $(\theta,\phi^I)$ in (\ref{NH4D5D}) to obtain charged scalar fields in AdS$_2$ with an electric field. From the $SL(2,R)$ transformation properties of these fields, one can read off the conformal weights $h_R$ of the dual operators.

Just as in 4d, operators dual to perturbations that break the rotational ($U(1) \times U(1)$) symmetry can have complex $h_R$. Therefore we focus on $U(1) \times U(1)$ invariant perturbations. Considering massless scalar field and electromagnetic perturbations as well as gravitational perturbations, we find the following spectrum
\begin{eqnarray}
 \textrm{massless scalar field perturbations: }&&h_R=\ell+1\ ,\label{5DsumS}\\
 \textrm{electromagnetic perturbations: }&&h_R=\ell+1,\ell+2,\ell+2\ ,\label{5DsumMX}\\
 \textrm{gravitational perturbations: }&&h_R=\ell+1,\ell+2,\ell+2,\ell+3,\ell+3\ ,\label{5DsumGR}
\end{eqnarray}
where $\ell=0,1,2,\cdots$.
We see that the conformal weights are integers again.\footnote{For the special case of near-horizon extreme Myers-Perry with equal angular momenta, this result was obtained in Ref. \cite{Durkee:2010ea}.}
This result seems more surprising than the 4d case because these 5d near-horizon geometries contain dimensionless parameters. Nevertheless, the conformal weights of these operators do not depend on these parameters. We do not have a microscopic interpretation of the results, but our results suggest that there is a "universal sector" present in all CFTs dual to extreme  rotating black holes.

Could this result be simply a consequence of symmetry? There are reasons to think otherwise. Ref \cite{Durkee:2010ea} studied perturbations of the near-horizon geometry of extreme cohomogeneity-1 Myers-Perry black holes (i.e. those with equal angular momenta in an odd number of dimensions). The analysis allowed for a negative cosmological constant. In the case of non-vanishing cosmological constant, $h_R$ is not longer integer-valued but various continuously with $r_+/\ell$ (where $r_+$ is the horizon radius and $\ell$ the AdS radius). Furthermore, even with vanishing cosmological constant, in more than five dimensions, it was found that $h_R$ was non-integer and could be complex, even in the rotationally invariant sector. This was interpreted as signalling a classical instability of the black hole of a type found earlier in Ref. \cite{Dias:2010eu}. In both of these examples, $h_R$ is not integer valued but the black hole is classically unstable (with a negative cosmological constant, there is a superradiant instability \cite{Kunduri:2006qa}). Only when the black hole is classically stable might we expect a dual CFT to exist, and only in this case does it happen that $h_R$ is integer valued. 

There are attempts to understand Kerr/CFT as a 
deformation of the CFT describing certain supersymmetric black
holes~\cite{Guica:2010ej,Compere:2010uk}. 
Other work has argued that extreme Myers-Perry and rotating Kaluza-Klein black holes can be
described by a bound state of D6- and D0-branes~\cite{Sheinblatt:1997nt,Itzhaki:1998ka,Emparan:2006it,Emparan:2007en}. It would be interesting to know if any of these approaches can be used to explain our result.

The organization of this paper is as follows.
In section~\ref{sec:GHP}, we explain the method to study  perturbations
 in near horizon geometries. In section~\ref{NHG5D},
we describe the 5d near horizon geometries of interest.
In sections~\ref{sec:scalar}, \ref{ConfEM} and \ref{ConfG}, 
conformal weights are studied
for operators dual to massless scalar field, electromagnetic and gravitational perturbations.
In section~\ref{stability}, we discuss the implications of our results for the classical stability of
the black holes considered here.

\section{Perturbations of near horizon geometries}
\label{sec:GHP}

Ref.~\cite{Durkee:2010ea} developed a formalism to study perturbations
of near horizon
geometries based on a decoupled equation derived in Ref.~\cite{Durkee:2010qu}. 
We briefly review the formalism in this section.

In the near horizon geometry~(\ref{NH4D5D}), 
the surface of constant $t$ and $r$ corresponds to a spatial 
cross-section of the black hole horizon.
We denote the surface by $\mathcal{H}$. 
The metric of $\mathcal{H}$ is 
\begin{equation}
 d\hat{s}^2=\hat{g}_{\mu\nu}dx^\mu dx^\nu=
\alpha(\theta)^2 d\theta^2
+\gamma_{IJ}(\theta)d\phi^I d\phi^J\ .
\end{equation}
We express gravitational and electromagnetic perturbations in terms of 
$\Omega_{ij}\equiv C_{abcd}\ell^a m^b_i \ell^c m^d_j$ and 
$\varphi_i\equiv F_{ab}\ell^a m_i^b$, respectively.
Here, $C_{abcd}$ is Weyl tensor and $F_{ab}$ is field strength of
electromagnetic field.
We introduced a null basis $\{\ell,n,m_i\}$ 
where $\ell=L(\theta)(-rdt+dr/r)/\sqrt{2}$, 
$n=L(\theta)(rdt+dr/r)/\sqrt{2}$
and $m_{i}$ are orthonormal spacelike vectors orthogonal to $\ell$ and $n$.
For scalar field equations $(\Box-M^2)\Phi=0$, we use the scalar field
$\Phi(t,r,\theta,\phi^I)$ as the perturbation variable.
For these perturbation variables, we assume separation anzatz as
\begin{equation}
\Phi=\chi_0(t,r)Y(\theta)\,e^{im_I\phi^I}\ ,\quad
\varphi_i=\chi_1(t,r)Y_{i}(\theta)\,e^{im_I\phi^I}\ ,\quad
\Omega_{ij}
=\chi_2(t,r)Y_{ij}(\theta)\,e^{im_I\phi^I}\ .
\end{equation}
where $m_I$ are constants (integers if $\phi^I$ have period $2\pi$). 
By  definition, $Y_{ij}$ is traceless and symmetric.
The perturbation equations reduce to equations
of massive charged Klein-Gordon equations in AdS$_2$ as
\begin{equation}
\label{2deq}
 [(\nabla_2-iq_s A_2)^2-q_s^2-\lambda_s]\chi_s=0\ ,\quad(s=0,1,2)
\end{equation}
where 
$\nabla_2$ is the covariant derivative on 
AdS$_2$ : $ds_2{}^2=-r^2dt^2+dr^2/r^2$ and
$A_2$ is a gauge field in AdS$_2$ given by $A_2=-rdt$.
The effective $U(1)$-charge $q_s$ is defined as $q_s=m_I k^I +is$.
The separation constant $\lambda_s$ is given by the eigenvalue equation
for angular directions, 
\begin{eqnarray}
&&(\mathcal{O}^{(0)}Y)=\lambda_0 Y\ ,\label{eigenS}\\
&&(\mathcal{O}^{(1)}Y)_{\mu}=\lambda_1 Y_{\mu}\ ,\label{eigenM}\\
&&(\mathcal{O}^{(2)}Y)_{\mu\nu}=\lambda_2 Y_{\mu\nu} \label{eigenG}\ ,
\end{eqnarray}
where these eigenvalue equations are written in coordinate basis $\{x^\mu\}=\{\theta,\phi^I\}$.
The operators $\mathcal{O}^{(s)}$ are defined as \cite{Durkee:2010ea}
\begin{eqnarray}
&&(\mathcal{O}^{(0)}Y)=-\hat{\nabla}_\mu(L(\theta)\hat{\nabla}^\mu Y)+L^2M^2-
(m_I k^I)^2\ ,\label{Oscalar}\\
&&(\mathcal{O}^{(1)}Y)_\mu
   =-\frac{1}{L^2}\hat{\nabla}^\rho (L^4 \hat{\nabla}_\rho Y_\mu )
      + \left(2-(k^I m_I)^2 - \frac{5}{4L^2}k_\mu k^\mu
     \right)Y_\mu \nonumber\\
      &&\hspace{3cm}+ L^2\left(\hat{R}_{\mu\nu} +
		   \frac{1}{2}\hat{R}\hat{g}_{\mu\nu}\right)Y^\nu
 + \bigg[-\frac{1}{2}(dk)_{\mu\nu} \nonumber\\
&&\hspace{6cm} + 2(k-d(L^2))_{[\mu} \hat{\nabla}_{\nu]}
            - \frac{1}{L^2} (dL^2)_{[\mu} k_{\nu]}\bigg] Y^\nu\ ,\label{Omaxwell}\\
  &&(\mathcal{O}^{(2)} Y)_{\mu\nu}
   =  -\frac{1}{L^4}\hat{\nabla}^\rho \left(L^6 \hat{\nabla}_\rho Y_{\mu\nu}\right)
      + \left(6-(k^I m_I)^2 - \frac{4}{L^2} k_\mu k^\mu  \right)Y_{\mu\nu} \nonumber\\
    &&\hspace{2cm}+ 2L^2\left(\hat{R}_{(\mu|\rho} 
             + \hat{R}\hat{g}_{(\mu|\rho}\right)Y^\rho_{\phantom{\rho}
    |\nu)}
      -
      2L^2\hat{R}_{\mu\phantom{\rho}\nu}^{\phantom{\mu}\rho\phantom{\nu}\sigma}Y_{\rho\sigma} \nonumber\\
&&\hspace{3cm}                                   
   + \Big[ - (dk)_{(\mu|\rho} - \frac{2}{L^2}\left(d(L^2)\wedge
					       k\right)_{(\mu|\rho} \nonumber\\
            &&\hspace{3cm}\qquad\qquad+ 2 \left( k-d(L^2)\right)_{(\mu|}\hat{\nabla}_\rho 
            - 2 \left( k-d(L^2)\right)_{\rho} \hat{\nabla}_{(\mu|}
            \Big]
                        Y^\rho_{\phantom{\rho} |\nu)}.
 \label{eqn:Ograv}
\end{eqnarray}
here, $\hat{\nabla}$, $\hat{R}_{\mu\nu\rho\sigma}$, $\hat{R}_{\mu\nu}$ and $\hat{R}$
are covariant derivative, Rieman tensor, Ricci tensor and Ricci scalar on
$\mathcal{H}$, respectively. 
We also defined $k=k^I\partial_I$ and $(dk)_{\mu\nu}=2\hat{\nabla}_{[\mu}k_{\nu]}$.

At large $r$, solutions of (\ref{2deq}) behave as 
$\chi_s\sim r^{-h_R},r^{h_R-1}$ where
\begin{equation}
 h_R=\frac{1}{2}+ \sqrt{\frac{1}{4}+\lambda_s}\ .
\label{expDelta}
\end{equation}
Assuming usual AdS/CFT rules apply, $h_R$ is the conformal weight of a dual CFT operator. Note that $q_s$ is complex in general but Ref.\cite{Durkee:2010ea} showed that the above operators are self-adjoint with respect to suitable inner products and hence $\lambda_s$ is real.

For perturbations with $m_I\neq 0$, $h_R$ can be complex even for Kerr black
holes~\cite{Bardeen:1999px,Dias:2009ex,Amsel:2009ev}. 
in this paper, we focus only on the
$U(1) \times U(1)$-invariant perturbations, i.e., we assume $m_I=0$.

\section{Near horizon geometries in 5-dimensions}
\label{NHG5D}

Assuming two $U(1)$ symmetries, 
one can classify the near horizon geometries of 5-dimensional extreme
vacuum black holes completely~\cite{Kunduri:2008rs,Hollands:2009ng}.
Any near horizon geometry is isometric to that of
Myers-Perry black hole, boosted Kerr black string or fast-rotating
Kaluza-Klein black hole.
In this section, we introduce these near horizon geometries.

\subsection{Myers-Perry black holes}

Using $L(\theta)$, $\alpha(\theta)$, $\gamma_{IJ}(\theta)$ and $k^I$
defined in Eq.(\ref{NH4D5D}),
we can describe the near horizon geometry of 
the extreme Myers-Perry black hole as~\cite{Kunduri:2007vf}
\begin{equation}
\begin{split}
&L^2(\theta)=\frac{\rho_+^2}{4}\ ,\quad
\alpha^2(\theta)=\rho_+^2\ ,\quad
k^{\phi^1}=-\frac{1}{2}\sqrt{\frac{b}{a}}\ ,\quad
k^{\phi^2}=-\frac{1}{2}\sqrt{\frac{a}{b}}\ ,\\
&\gamma_{IJ}(\theta)d\phi^Id\phi^J=\frac{(a+b)^2}{\rho_+^2}
\bigg[
a\sin^2\theta(a+b\sin^2\theta) (d\phi^1)^2\\
&\hspace{4cm}+b\cos^2\theta(b+a\cos^2\theta) (d\phi^2)^2
+2ab\sin^2\theta\cos^2\theta d\phi^1 d\phi^2
\bigg]\ ,
\end{split}
\label{NHMP2}
\end{equation}
and $\rho_+^2=ab+a^2\cos^2\theta+b^2\sin^2\theta$. 
The coordinate ranges are $0\leq\theta\leq\pi/2$ and $0\leq\phi^I\leq 2\pi$.
This spacetime is parametrized by two parameters $a$ and $b$, which represent
angular momentum parameters in the full geometry. 

\subsection{Boosted Kerr-strings}

\label{sec:string}

The near horizon data of an extreme boosted Kerr-string has the form (\ref{NH4D5D}) with $\phi^I \sim (\phi,z)$ and \cite{Kunduri:2007vf}
\begin{equation}
\begin{split}
&L^2(\theta)=\alpha^2(\theta)=2a^2\Omega^2(\theta)\ ,\quad
k^\phi=-1\ ,\quad k^z=0\ ,\\
&\gamma_{IJ}(\theta)d\phi^Id\phi^J
=2a^2\Omega^2(\theta)\Lambda^2(\theta)\left(d\phi+\frac{\sinh\beta}{2a}dz\right)^2
+\cosh^2\beta dz^2
\end{split}
\label{NHEKS}
\end{equation}
where functions $\Omega(\theta)$ and $\Lambda(\theta)$ are defined as
\begin{equation}
 \Omega^2(\theta)=\frac{1+\cos^2\theta}{2}\ ,\quad
\Lambda(\theta)=\frac{2\sin\theta}{1+\cos^2\theta}\ .
\end{equation}
The coordinate ranges are $0\leq\theta\leq\pi$, $0\leq\phi\leq 2\pi$ and
$-\Delta z/2 \leq z \leq\Delta z/2$.
This spacetime is parametrized by three parameters $a$, $\beta$ and 
$\Delta z$. The near horizon geometries of extreme doubly spinning black rings~\cite{Pomeransky:2006bd} are
included in Eq.(\ref{NHEKS}) as a special case corresponding to 
\begin{equation}
 a^2=\frac{8\ell^2\lambda^2}{(4-\lambda^2)^2}\ ,\quad
 \sinh^2\beta=1\ ,\quad
 \Delta z=\frac{2\sqrt{2}\pi \ell(2+\lambda)}{2-\lambda}\ ,
\label{DSBRpara}
\end{equation}
where $0\leq\lambda<2$ and $\ell>0$. 

For the $U(1)\times U(1)$ invariant perturbations that we shall consider, the absence of any $\phi$ or $z$ dependence in the perturbation means that there will be no dependence of our equations on $\Delta z$. Hence the equations will be the same as for $\Delta z = \infty$. But in this limit, the parameter $\beta$ can be eliminated by a shift in $\phi$ and a rescaling of $z$. This reduces the metric to that of an unboosted extreme Kerr string with a single parameter $a$, which is dimensionful and therefore cannot affect $h_R$. This explains why, for this case, $h_R$ does not depend on the parameters of the solution.

\subsection{Fast-rotating Kaluza-Klein black holes}

We consider rotating Kaluza-Klein black hole
derived in~\cite{Rasheed:1995zv,Larsen:1999pp}.
The black hole has spherical horizon and approaches
the Gross-Perry-Sorkin  monopole solution~\cite{Gross:1983hb,Sorkin:1983ns} at infinity.
The non-extremal solution has four parameters, $(m,a,p,q)$ which
correspond to mass, angular momenta, electric charges and magnetic
charge in effective 4-dimensional theory.
In this spacetime, there are two kind of extreme limit.
One of them is $a,m\to 0$ with $a/m$ fixed.
In this limit, the solution is called the slow-rotating extreme
Kaluza-Klein black hole. The near horizon geometry of this spacetime is
isometric to that of extreme Myers-Perry black hole.
Another extreme limit is $m\to a$.
In this limit, the solution is called the fast-rotating extreme
Kaluza-Klein black hole. We consider the near horizon geometry of the
fast-rotating extreme KKBH.

The near horizon data of the fast-rotating KKBH has the form (\ref{NH4D5D}) with $\phi^I \sim (y,\phi)$ and~\cite{Kunduri:2008rs}
\begin{equation}
\begin{split}
 &L^2(\theta)=\alpha^2(\theta)=H_p\ ,\quad
 k^\phi=\frac{pq+4a^2}{2a(p+q)}\ ,\quad
 k^y=-\frac{\sqrt{pq}(p^2-4a^2)Q}{qa(p+q)}\\
 &\gamma_{IJ}(\theta) d\phi^I d\phi^J=
 \frac{H_q}{H_p}(dy+A_\phi d\phi)^2 
+\frac{pqa^2\sin^2\theta}{H_q}d\phi^2\ ,
\end{split}
\label{fastKKBG}
\end{equation}
where
\begin{equation}
\begin{split}
&H_p=-a^2\sin^2\theta+\frac{p(pq+4a^2)}{2(p+q)}+\frac{2pQP}{\sqrt{pq}}\cos\theta\ ,\\
&H_q=-a^2\sin^2\theta+\frac{q(pq+4a^2)}{2(p+q)}-\frac{2qQP}{\sqrt{pq}}\cos\theta\
 ,\\
&A_\phi=-\frac{2P}{H_q}(H_q+a^2\sin^2\theta)\cos\theta
+\sqrt{\frac{p}{q}}\frac{Q(2a^2(p+q)+q(p^2-4a^2))\sin^2\theta}{(p+q)H_q}\ ,
\end{split}
\end{equation}
and 
\begin{equation}
P=\sqrt{\frac{p(p^2-4a^2)}{4(p+q)}}\ ,\quad
Q=\sqrt{\frac{q(q^2-4a^2)}{4(p+q)}}\ .
\end{equation}
The coordinate ranges are $0\leq\theta\leq\pi$, $0\leq\phi\leq 2\pi$ and
$0\leq y \leq 8\pi P$.
This solution is parametrized by three parameters, $(p,q,a)$ which satisfy
$p,q\geq 2a$.

\section{Conformal weights for massless scalar field perturbations}
\label{sec:scalar}

In sections~\ref{sec:scalar}, \ref{ConfEM} and \ref{ConfG}, 
we solve the eigenvalue equations~(\ref{eigenS}-\ref{eigenG}) and determine conformal
dimensions of CFT operators dual to $U(1) \times U(1)$-invariant perturbations of the 5d near horizon geometries
introduced in section \ref{NHG5D}.
In this section, we consider scalar field perturbations~(\ref{eigenS}).
For the general near horizon geometry~(\ref{NH4D5D}), 
the eigenvalue equation for the scalar field  reduces to
\begin{equation}
  -\frac{L^2}{\alpha^2} Y''
-\frac{1}{2L^2 \gamma}\left(\frac{L^4\gamma}{\alpha^2}\right)'
Y'+L^2\{M^2-m_Im_J (\gamma^{IJ}+k^Ik^J)\}=\lambda_0 Y\ ,
\label{scalar}
\end{equation}
where $'=d/d\theta$ and $\gamma=\textrm{det}(\gamma_{IJ})$. Hereafter, we consider the massless case $M=0$ and
$U(1) \times U(1)$-invariant perturbations, i.e., $m_I=0$.  
For extreme Myers-Perry black holes, Kerr-strings, and 
fast-rotating Kaluza-Klein black holes, we find the relation
\begin{equation}
\begin{split}
&\frac{L^2}{\alpha^2}=\frac{1}{4}\ ,\quad L^2 \gamma =
 \frac{1}{16}ab\,(a+b)^4\sin^2 2\theta \quad(\textrm{Myers-Perry BH})\\
&\frac{L^2}{\alpha^2}=1\ ,\quad 
L^2 \gamma =4a^4\cosh^2\beta
 \sin^2\theta \quad(\textrm{Kerr-string})\ ,\\
&\frac{L^2}{\alpha^2}=1\ ,\quad 
L^2 \gamma = 16pq a^2 P^2\sin^2\theta \quad(\textrm{fast-rotating KKBH})\ .
\end{split}
\end{equation}
After a coordinate transformation $2\theta\to\theta$ for the Myers-Perry
black hole, 
substituting these functions into (\ref{scalar}), 
we obtain same
eigenvalue equation for three backgrounds as
\begin{equation}
 -Y''-\cot\theta Y'=\lambda_0 Y\ .
\end{equation}
The solution is expressed by Legendre functions as
\begin{equation}
 Y=C_1\,P_{\nu}{}(\cos\theta)+C_2\,Q_{\nu}{}(\cos\theta)\
  ,
\quad(\nu=\sqrt{\lambda_0+1/4}-1/2)
\end{equation}
The regularity requires $C_2=0$ and 
$\lambda_0 = \ell(\ell+1)$ $(\ell=0,1,2,\cdots)$.
Thus, using Eq.(\ref{expDelta}), we can determine conformal weights as
\begin{equation}
 h_R=\ell+1\ ,\quad(\ell=0,1,2,\cdots)\ .
\end{equation}
The conformal weights are integers and do not
depend on the parameters of the near-horizon geometry.

\section{Conformal weights for electromagnetic perturbations}
\label{ConfEM}

\subsection{Myers-Perry black holes}
Now we study the conformal weights of CFT operators dual to
electromagnetic perturbations~(\ref{eigenM}).
In this subsection, we consider the near horizon geometry of the extreme Myers-Perry black
hole~(\ref{NHMP2}).
For later numerical calculations, we define a set of variables
$\{\psi_1,\psi_2,\psi_3\}$ as
\begin{equation}
\begin{split}
&Y_\theta=\psi_1(\theta)\sin\theta\cos\theta\ ,\\
&Y_{\phi_1}=\psi_2(\theta)\sin^2\theta\{\sin^2\theta+(a/b)\cos^4\theta\}+\sin^4\theta\cos^2\theta\psi_3(\theta)\
 ,\\
&Y_{\phi_2}=\psi_2(\theta)\sin^4\theta\cos^2\theta+\psi_3(\theta)\cos^2\theta\{\cos^2\theta+(b/a)\sin^4\theta\}
\end{split}
\label{redef_MXW}
\end{equation}
Substituting Eq.(\ref{redef_MXW}) into, Eq.(\ref{eigenM}), we
obtain the eigenvalue equation for $\psi_k$ as
\begin{equation}
 \left[\frac{1}{4}\partial_\theta^2 + A(\theta;a/b) \partial_\theta
  +B(\theta;a/b)\right]\vec{\psi}=-\lambda_1 \vec{\psi}\ .
\label{eigen2}
\end{equation}
where $\vec{\psi}=(\psi_1,\psi_2,\psi_3)^T$ and
$A(\theta;a/b)$ and $B(\theta;a/b)$ are $3\times 3$ matrices which
depend on $\theta$ and $a/b$. Since $A$ and $B$ are dimensionless,
these can depend only on dimensionless parameter $a/b$.\footnote{
Because explicit expressions of $A$ and $B$ are tedious,
we do not write the expressions in this paper.}
In the case of equal angular momenta $a=b$, 
the symmetry of the near horizon spacetime enhances and 
we can solve the eigenvalue
equation~(\ref{eigen2}) analytically~\cite{Durkee:2010ea}.
Then, Eq.(\ref{expDelta}) gives the conformal weights as
\begin{equation}
\begin{split}
 h_R&=2,\kappa,\kappa+1,\kappa+2\,(\kappa=1,2,3,\cdots)\\
&=1,2,2,2,3,3,3,4,4,4,5,5,5,\cdots\ .
\end{split}
\label{a=b_MXW}
\end{equation}
For $a\neq b$, we need numerical calculations.
Thanks to the definition of $\psi_k$~(\ref{redef_MXW}), the regular solution behaves as 
$\vec{\psi}\sim \theta^0$ ($\theta\to 0$) and 
$\vec{\psi}\sim (\pi/2-\theta)^0$ ($\theta\to \pi/2$) at axis.
Hence, we simply impose the Neumann boundary conditions at
$\theta=0,\pi/2$ in our numerical calculations.
The detail of numerical calculations are summarized in
Appendix \ref{app.num}.
We show the numerical result 
 in Table \ref{MaxMPBH}.
We can find that the conformal weights do not depend on the parameter
$a/b$. Thus, our numerical calculations indicate that the
expression~(\ref{a=b_MXW}) is valid even for $a\neq b$. It would be nice to prove this analytically.

\begin{table}[h]
{\footnotesize{
\begin{equation}
\begin{array}{ c| c c c c c}
& a=b & a=2b & a=5b &a=10b &a=20b \\ \hline
1 & 1.000000 & 0.999999 & 0.999999 & 1.000000 & 1.000002 \\
2 & 1.999999 & 1.999998 & 1.999997 & 1.999997 & 1.999998 \\
3 & 1.999999 & 1.999998 & 1.999998 & 1.999998 & 2.000000 \\
4 & 1.999999 & 1.999999 & 1.999999 & 2.000000 & 2.000004 \\
5 & 2.999992 & 2.999991 & 2.999990 & 2.999989 & 2.999989 \\
6 & 2.999993 & 2.999992 & 2.999992 & 2.999992 & 2.999995 \\
7 & 2.999993 & 2.999993 & 2.999993 & 2.999995 & 3.000000 \\
8 & 3.999976 & 3.999976 & 3.999975 & 3.999972 & 3.999966 \\
9 & 3.999978 & 3.999977 & 3.999976 & 3.999975 & 3.999974 \\
10 & 3.999978 & 3.999978 & 3.999977 & 3.999981 & 3.999993 \\
11 & 4.999948 & 4.999948 & 4.999944 & 4.999940 & 4.999936 \\
12 & 4.999950 & 4.999948 & 4.999949 & 4.999948 & 4.999947 \\
13 & 4.999950 & 4.999951 & 4.999950 & 4.999952 & 4.999959 \\
\end{array}
\nonumber
\end{equation}
}}
\caption{\label{MaxMPBH} 
Conformal weights for operators dual to $U(1) \times U(1)$-invariant electromagnetic perturbations on near horizon
 geometries of extreme Myers-Perry black holes.
They are computed for $a/b=1,2,5,10$ and $20$.
We see that the conformal weights do not depend on the dimensionless
 parameter $a/b$ and the sequences of conformal weights coincide with
Eq.(\ref{a=b_MXW}) even for $a\neq b$. 
}
\end{table}

\subsection{Boosted Kerr-strings}

In this subsection, we consider the electromagnetic perturbations on
near horizon geometries of boosted Kerr-strings~(\ref{NHEKS}).
We define orthogonal basis on $\mathcal{H}$ as
\begin{equation}
\begin{split}
 &m^{(1)}_\mu dx^\mu
=ad\theta+a\Lambda(\theta)\cos\theta\left(d\phi+\frac{\sinh\beta}{2a}dz\right)
\ ,\\
 &m^{(2)}_\mu dx^\mu
=-a\cos\theta
 d\theta+a\Lambda(\theta)\left(d\phi+\frac{\sinh\beta}{2a}dz\right)\ ,\quad
 m^{(3)}_\mu dx^\mu
=\cosh\beta dz\ .
\end{split}
\label{basis_KS}
\end{equation}
and rewrite the $Y_{\mu}$ in this basis as 
$Y_{\mu}=Y_{i}m^i_\mu$. 
We define a set of variables $\psi_k$ ($k=1,\cdots,3$) as
\begin{equation}
 Y_1=\psi_2/\sqrt{1+\cos^2\theta}\ ,\quad
 Y_2=\psi_3/\sqrt{1+\cos^2\theta}\ ,\quad
 Y_3=\psi_1/\sqrt{1+\cos^2\theta}
\end{equation}
In term of $\psi$'s, we obtain five eigenvalue equations 
decoupled each other,
\begin{equation}
  \psi_k''+\cot\theta \psi_k' -\frac{s_k^2}{\sin^2\theta}\psi_k= -\lambda_1
  \psi_k\ ,\quad(k=1,2,3)
\end{equation}
where $(s_1,s_2,s_3)=(0,1,1)$. 
Considering the Kaluza-Klein reduction along the $z$ direction in~(\ref{NHEKS}),
we can regard the variable $\{\psi_1\}$ and $\{\psi_2,\psi_3\}$
as effective scalar and vector field in NHEK
geometry, respectively.
These equations for $\psi_k$ do not depend on parameters $a$ and $\beta$. 
The solutions are expressed by associated Legendre functions as
\begin{equation}
 \psi_k=C_1\,P_{\nu}{}^{s_k}(\cos\theta)+C_2\,Q_{\nu}{}^{s_k}(\cos\theta)\
  ,
\quad(\nu=\sqrt{\lambda_1+1/4}-1/2)
\end{equation}
The regularity of the solution requires $C_2=0$ and 
$\lambda_1 = (\ell+s_k)(\ell+s_k+1)$ $(\ell=0,1,2,\cdots)$.
Thus, conformal weights are 
\begin{equation}
\begin{split}
 h_R&=\ell+1,\ell+2,\ell+2 \quad (\ell=0,1,2,\cdots)\\
&=1,2,2,2,3,3,3,4,4,4,5,5,5,\cdots\ .
\end{split}
\label{NHEKstring_MW}
\end{equation}
Hence extreme Myers-Perry black holes and Kerr black strings have
same conformal weights for operators dual to $U(1) \times U(1)$-invariant electromagnetic perturbations.

\subsection{Fast-rotating Kaluza-Klein black holes}

In this subsection, we consider the electromagnetic perturbations on
near horizon geometries of fast-rotating Kaluza-Klein black
holes~(\ref{fastKKBG}).
We define the basis on $\mathcal{H}$ as
\begin{equation}
 m^{(1)}=d\theta\ ,\quad m^{(2)}=dy+A_\phi d\phi\ ,\quad
 m^{(3)}=d\phi\ ,
\label{basis_fKK}
\end{equation}
We rewrite the $Y_{\mu}$ in this basis as 
$Y_{\mu}=Y_{i}m^i_\mu$ and
define variables $\psi_k$ ($k=1,\cdots,3$) as
\begin{equation}
\begin{split}
&Y_1=\psi_1(\theta)\sin\theta\ ,\\
&Y_2=\psi_2(\theta)(\beta_+\cos^2(\theta/2)+\beta_-\sin^2(\theta/2))\sin^2\theta+\psi_3(\theta)\\
&Y_3=\psi_3(\theta)\sin^2\theta
\end{split}
\end{equation}
where $\beta_{\pm}$ are constants defined by
\begin{equation}
\beta_{\pm}=
\frac{
\{
(pq)^{1/2}(
pq+4{a}^{2})
\mp 4(p+q)QP
\}
\{
(pq)^{1/2}P(
{q}^2-4{a}^{2})
\pm pQ(pq
+4{a}^{2})\}
 }
{2p^{3/2}q^{1/2}(p+q)
\{
\pm (pq)^{1/2}(pq+4a^2)
+4(p+q)QP\}
 {a}^{2}}
\end{equation}
Then, we obtain the eigenvalue equations for $\psi_k$ as
\begin{equation}
  \left[\partial_\theta^2 + C(\theta;p/a,q/a) \partial_\theta
  +D(\theta;p/a,q/a)\right]\vec{\psi}=-\lambda_1 \vec{\psi}\ .
\label{eigen3}
\end{equation}
where $\vec{\psi}=(\psi_1,\psi_2,\psi_3)^T$ and
$C$ and $D$ are $3\times 3$ matrices which
depend on $\theta$ and $p/a$ and $q/a$.
We solved the equations numerically. 
Since the regular solution behaves as 
$\vec{\psi}\sim \theta^0$ ($\theta\to 0$) and 
$\vec{\psi}\sim (\pi-\theta)^0$ ($\theta\to \pi$) at axis, we
imposed Neumann boundary conditions at axis.
We summarize the conformal weights for several parameters in
Table \ref{fKKMXW}.
We see that the conformal weights do not depend on the parameters, $(p/a,q/a)$ (it would be nice to prove this analytically).
Furthermore, the sequence of conformal weights is same as that of
Myers-Perry black holes~(\ref{a=b_MXW}) and Kerr-strings~(\ref{NHEKstring_MW}).
From these results, we can conclude that CFT operators dual to $U(1) \times U(1)$-invariant
electromagnetic perturbations on 5d near horizon geometries
have same integer conformal weights~(\ref{5DsumMX}).

\begin{table}[h]
{\footnotesize{
\begin{equation}
\begin{array}{ c| c c c c}
& (3,3) & (3,6) & (6,3) & (6,6) \\ \hline
1 & 1.000000 & 1.000000 & 1.000000 & 0.999999 \\
2 & 1.999998 & 1.999998 & 1.999998 & 1.999997 \\
3 & 1.999999 & 1.999998 & 1.999998 & 1.999998 \\
4 & 1.999999 & 1.999999 & 1.999999 & 1.999998 \\
5 & 2.999992 & 2.999991 & 2.999991 & 2.999991 \\
6 & 2.999992 & 2.999992 & 2.999992 & 2.999991 \\
7 & 2.999993 & 2.999992 & 2.999992 & 2.999991 \\
8 & 3.999975 & 3.999975 & 3.999976 & 3.999974 \\
9 & 3.999978 & 3.999976 & 3.999977 & 3.999976 \\
10 & 3.999978 & 3.999977 & 3.999977 & 3.999977 \\
11 & 4.999948 & 4.999948 & 4.999947 & 4.999945 \\
12 & 4.999950 & 4.999948 & 4.999949 & 4.999948 \\
13 & 4.999950 & 4.999950 & 4.999950 & 4.999949 \\
\end{array}
\nonumber
\end{equation}
}}
\caption{\label{fKKMXW} 
Conformal weights for operators dual to $U(1) \times U(1)$-invariant electromagnetic perturbations on near horizon
 geometries of fast-rotating Kaluza-Klein black holes.
They are computed for $(p/a,q/a)=(3,3)$, $(3,6)$, $(6,3)$ and $(6,6)$.
We see that the conformal weights are integer and do not depend on
 dimensionless parameters, $p/a$ and $q/a$.
}
\end{table}

\section{Conformal weights for gravitational perturbations}
\label{ConfG}

\subsection{Myers-Perry black holes}

Finally, we consider the gravitational perturbations on 5-dimensional
near horizon geometries.
In this subsection, we consider Myers-Perry black holes.
We define variables $\psi_k$ ($k=1,\cdots,5$) as
\begin{equation}
\begin{split}
&Y_{\theta\phi^1}=\psi_1(\theta)\sin^3\theta\cos\theta\ ,\quad
Y_{\theta\phi^2}=\psi_2(\theta)\sin\theta\cos^3\theta\ ,\quad
Y_{\phi^1\phi^2}=\psi_3(\theta)\sin^2\theta\cos^2\theta\ ,\\
&Y_{\phi^1\phi^1}=\psi_4(\theta)(1+\cos^2\theta)\sin^2\theta -\frac{a^2}{2b(a+b)}\psi_5(\theta)\cos^2\theta\ ,\\
&Y_{\phi^2\phi^2}=\psi_5(\theta)(1+\sin^2\theta)\cos^2\theta -\frac{b^2}{2a(a+b)}\psi_4(\theta)\sin^2\theta\ .
\end{split}
\label{redef}
\end{equation}
The $Y_{\theta\theta}$ component is determined by the traceless
condition. 
The regular solution behaves  at axis as 
$\vec{\psi}\sim \theta^0$ ($\theta\to 0$) and 
$\vec{\psi}\sim (\pi/2-\theta)^0$ ($\theta\to \pi/2$).
In the case of equal angular momenta $a=b$,
the eigenvalue equation~(\ref{eigenG}) has been solved analytically~\cite{Durkee:2010ea}.
Then, conformal weights are
\begin{equation}
\begin{split}
 h_R&=2,3,3,4,\kappa-1,\kappa,\kappa+1,\kappa+2,\kappa+3,\quad(\kappa=2,3,4,\cdots)\\
&=1,2,2,2,3,3,3,3,3,4,4,4,4,4,5,5,5,5,5,\cdots\ .
\end{split}
\label{a=b_grav}
\end{equation}
For $a\neq b$, we solved the eigenvalue equation numerically by the same
way as Eq.(\ref{eigen2}) and (\ref{eigen3}).
The result is summarized in Table \ref{del}.
We can find that the conformal weights do not depend on the parameter
$a/b$ again. Thus, our numerical calculations indicate that the
expression~(\ref{a=b_grav}) is valid even for $a\neq b$.
\begin{table}[h]
{\footnotesize{
\begin{equation}
\begin{array}{ c| c c c c c}
& a=b & a=2b & a=5b &a=10b &a=20b \\ \hline
1 & 1.000001 & 0.999999 & 0.999993 & 0.999990 & 0.999997 \\
2 & 1.999999 & 1.999998 & 1.999994 & 1.999980 & 1.999969 \\
3 & 1.999999 & 1.999998 & 1.999997 & 2.000001 & 2.000010 \\
4 & 1.999999 & 1.999998 & 1.999998 & 2.000005 & 2.000016 \\
5 & 2.999992 & 2.999991 & 2.999985 & 2.999986 & 2.999975 \\
6 & 2.999993 & 2.999993 & 2.999992 & 2.999988 & 2.999993 \\
7 & 2.999993 & 2.999993 & 2.999992 & 2.999991 & 2.999996 \\
8 & 2.999993 & 2.999993 & 2.999993 & 2.999992 & 2.999997 \\
9 & 2.999993 & 2.999993 & 2.999993 & 2.999995 & 2.999998 \\
10 & 3.999977 & 3.999975 & 3.999968 & 3.999970 & 3.999952 \\
11 & 3.999978 & 3.999978 & 3.999977 & 3.999972 & 3.999975 \\
12 & 3.999978 & 3.999978 & 3.999978 & 3.999975 & 3.999980 \\
13 & 3.999979 & 3.999979 & 3.999978 & 3.999978 & 3.999985 \\
14 & 3.999979 & 3.999979 & 3.999980 & 3.999980 & 3.999990 \\
15 & 4.999950 & 4.999948 & 4.999941 & 4.999929 & 4.999912 \\
16 & 4.999951 & 4.999950 & 4.999949 & 4.999944 & 4.999952 \\
17 & 4.999951 & 4.999951 & 4.999950 & 4.999949 & 4.999954 \\
18 & 4.999952 & 4.999952 & 4.999950 & 4.999956 & 4.999959 \\
19 & 4.999953 & 4.999953 & 4.999954 & 4.999958 & 4.999961 \\
\end{array}
\nonumber
\end{equation}
}}
\caption{\label{del} 
Conformal weights for operators dual to $U(1) \times U(1)$-invariant gravitational perturbations on near horizon
 geometries of extreme Myers-Perry black holes.
They are computed for $a/b=1,2,5,10$ and $20$.
We see that the conformal weights do not depend on the dimensionless
 parameter $a/b$ and the sequences of conformal weights coincide with
Eq.(\ref{a=b_grav}) even for $a\neq b$. 
}
\end{table}

\subsection{Boosted Kerr-strings}

In this subsection, we consider Boosted Kerr-strings.
We use the orthogonal basis defined in Eq.(\ref{basis_KS}).
We define a set of variables $\psi_k$ ($k=1,\cdots,5$) as
\begin{equation}
\begin{split}
& Y_{12}=\psi_4(\theta)/(1+\cos^2\theta)\ ,\quad
 Y_{13}=\psi_2(\theta)/(1+\cos^2\theta)\ ,\\
& Y_{22}=(\psi_5(\theta)-\psi_1(\theta))/(1+\cos^2\theta)\ ,\quad
 Y_{23}=\psi_3(\theta)/(1+\cos^2\theta)\ ,\\
& Y_{33}=2\psi_1(\theta)/(1+\cos^2\theta)\ .
\end{split}
\end{equation}
In term of $\psi$'s, we obtain five eigenvalue equations 
decoupled each other,
\begin{equation}
 \psi_k''+\cot\theta \psi_k' -\frac{s_k^2}{\sin^2\theta}\psi_k= -\lambda_2
  \psi_k\ ,
\end{equation}
where $(s_1,s_2,s_3,s_4,s_5)=(0,1,1,2,2)$. 
After the Kaluza-Klein reduction along the $z$ direction,
the variable $\{\psi_1\}$, $\{\psi_2,\psi_3\}$ and
$\{\psi_4,\psi_5\}$  correspond to effective scalar, vector and tensor field in NHEK
geometry, respectively.
These equations for $\psi_k$ do not depend on parameters $a$ and $\beta$. 
The solutions are expressed by associated Legendre functions as
\begin{equation}
 \psi_k=C_1\,P_{\nu}{}^{s_k}(\cos\theta)+C_2\,Q_{\nu}{}^{s_k}(\cos\theta)\
  ,
\quad(\nu=\sqrt{\lambda_2+1/4}-1/2)
\end{equation}
The regularity of the solution requires $C_2=0$ and 
$\lambda_2 = (\ell+s_k)(\ell+s_k+1)$ $(\ell=0,1,2,\cdots)$.
Thus, the conformal weights are 
\begin{equation}
\begin{split}
 h_R&=\ell+1,\ell+2,\ell+2,\ell+3,\ell+3 \quad (\ell=0,1,2,\cdots)\\
&=1,2,2,2,3,3,3,3,3,4,4,4,4,4,5,5,5,5,5,\cdots\ .
\end{split}
\end{equation}
This is precisely what one expects on the basis of the remarks at the end of section \ref{sec:string}: the spectrum should coincide with that of an unboosted Kerr string. This can be understood from Kaluza-Klein theory: in 4d one has massless scalar, vector and gravitational perturbations of the Kerr black hole, for which the conformal weights can be read off from Refs. \cite{Bardeen:1999px,Dias:2009ex,Amsel:2009ev}.

In summary, extreme Myers-Perry black holes and Kerr black strings have
give the same conformal weights for operators dual to $U(1)\times U(1)$-symmetric gravitational perturbations.

\subsection{Fast-rotating Kaluza-Klein black holes}

We use the orthogonal basis defined in Eq.(\ref{basis_fKK}).
We define variables $\psi_k$ as
\begin{equation}
\begin{split}
&Y_{12}=\psi_1(\theta)\sin\theta\ ,\quad
Y_{13}=\psi_2(\theta)\sin^3\theta\ ,\\
&Y_{22}=(\alpha_+\cos^2(\theta/2)+\alpha_-\sin^2(\theta/2))\psi_3(\theta)\
 ,\\
&Y_{23}=\psi_4(\theta)\sin^2\theta\ ,\quad
Y_{33}=(\psi_3(\theta)+\psi_5(\theta)\sin^2\theta)\sin^2\theta\ ,
\end{split}
\end{equation}
where
\begin{equation}
 \alpha_\pm=
-\frac{\{(pq)^{3/2}+4(pq)^{1/2}a^2 \mp 4(p+q)PQ\}^2(pq)^{1/2}}
{p^3(p+q)a^2\{(pq)^{3/2}+4(pq)^{1/2}a^2 \pm 4(p+q)PQ\}}\ .
\end{equation}
The regular solution behaves as 
$\vec{\psi}\sim \theta^0$ ($\theta\to 0$) and 
$\vec{\psi}\sim (\pi-\theta)^0$ ($\theta\to \pi$).
We solved the eigenvalue equation numerically and 
summarize the conformal weights for several parameters in
Table \ref{del2}.
We see that the conformal weights do not depend on the dimensionless parameters, $(p/a,q/a)$.
Furthermore, the sequence of conformal weights is same as that of
Myers-Perry black holes~(\ref{a=b_grav}) and Kerr-strings.
Form these results, we can conclude that CFT operators dual to $U(1) \times U(1)$-symmetric gravitational perturbations of any 5d vacuum near horizon geometry have the same integer conformal weights~(\ref{5DsumGR}).
\begin{table}[h]
{\footnotesize{
\begin{equation}
\begin{array}{ c| c c c c}
& (3,3) & (3,6) & (6,3) & (6,6) \\ \hline
1 & 1.000000 & 0.999994 & 0.999993 & 1.000057 \\
2 & 1.999998 & 1.999995 & 1.999993 & 1.999995 \\
3 & 1.999999 & 1.999996 & 1.999997 & 2.000018 \\
4 & 1.999999 & 1.999998 & 1.999999 & 2.000038 \\
5 & 2.999991 & 2.999987 & 2.999986 & 2.999989 \\
6 & 2.999993 & 2.999990 & 2.999991 & 2.999990 \\
7 & 2.999993 & 2.999991 & 2.999992 & 2.999991 \\
8 & 2.999993 & 2.999992 & 2.999992 & 3.000005 \\
9 & 2.999994 & 2.999993 & 2.999992 & 3.000039 \\
10 & 3.999977 & 3.999969 & 3.999972 & 3.999974 \\
11 & 3.999978 & 3.999976 & 3.999976 & 3.999977 \\
12 & 3.999978 & 3.999978 & 3.999977 & 3.999981 \\
13 & 3.999978 & 3.999978 & 3.999978 & 3.999986 \\
14 & 3.999980 & 3.999978 & 3.999978 & 4.000020 \\
15 & 4.999950 & 4.999939 & 4.999943 & 4.999944 \\
16 & 4.999950 & 4.999951 & 4.999950 & 4.999949 \\
17 & 4.999951 & 4.999951 & 4.999950 & 4.999949 \\
18 & 4.999952 & 4.999951 & 4.999951 & 4.999952 \\
19 & 4.999954 & 4.999951 & 4.999951 & 5.000000 \\
\end{array}
\nonumber
\end{equation}
}}
\caption{\label{del2} 
Conformal weights for gravitational perturbations on near horizon
 geometries of fast-rotating Kaluza-Klein black holes.
They are computed for $(p/a,q/a)=(3,3)$, $(3,6)$, $(6,3)$ and $(6,6)$.
We see that the conformal weights are integer and do not depend on
 dimensionless parameters, $p/a$ and $q/a$.
}
\end{table}

\section{Classical stability of extreme black holes}
\label{stability}

In~\cite{Durkee:2010ea}, it was conjectured that, if rotationally invariant perturbations of the near
horizon geometry of an extreme black holes lead to complex $h_R$ then the full black hole spacetime is classically unstable. As examples, they studied the instability of odd dimensional
Myers-Perry black holes with equal angular momenta and had consistent
results with the stability analysis of full black hole geometries near to extremality~\cite{Murata:2008yx,Dias:2010eu}. 
In our notation, the conjecture applies to $U(1)\times U(1)$-invariant perturbations and the condition for instability is given by
$\lambda_s<-1/4$. ($\lambda_s$ was defined in Eq.(\ref{eigenS}-\ref{eigenG}).)
From our result~(\ref{5DsumGR}), we find $h_R \geq 1$ so
$\lambda_s\geq 0$. Therefore, the conjecture of Ref.~\cite{Durkee:2010ea} does not predict any instability for axisymmetric perturbations of 5-dimensional extreme black holes.

This is consistent with expectations about stability of the full 5d Myers-Perry geometry.
It has been shown that 5-dimensional Schwarzschild black holes are
stable~\cite{Kodama:2003jz,Ishibashi:2003ap}.
Hence, if 5-dimensional Myers-Perry black holes are unstable for axisymmetric
perturbations, the stability must change at a critical value
of the angular momentum. Previous studies of axisymmetric instabilities have found that there exists a time-independent perturbation at this critical value of the angular momentum \cite{Dias:2009iu,Dias:2010eu} which is expected to indicate the bifurcation of a new family of stationary black hole solutions. On the other hand, however, it has been shown that 
the Myers-Perry solution is the unique
5d stationary, asymptotically flat, solution with two commuting $U(1)$ symmetries and 
topologically spherical horizon~\cite{Morisawa:2004tc,Hollands:2007aj}.
Thus, there cannot be unstable modes in $U(1) \times U(1)$-invariant perturbation of 
5-dimensional Myers-Perry black holes.
Our result~(\ref{5DsumGR}) is consistent with this notion and supports
the conjecture in~\cite{Durkee:2010ea}.

 \section*{Acknowledgments}
We would like to thank T.~Nishioka, J.~Soda and Y.~Yasui for
discussions and especially to H.~Reall for collaboration at an earlier stage.
KM is supported by a grant for research abroad by the JSPS (Japan).

\appendix

\section{Numerical calculations}
\label{app.num}

In this section, we explain how to solve Eq.(\ref{eigen2}),
numerically. We carried out other numerical calculations in this paper 
by the same way as Eq.(\ref{eigen2}).

Instead of Eq.(\ref{eigen2}), we consider a diffusion equation,
\begin{equation}
 \partial_\tau\vec{\psi}(\tau,\theta)=
M\vec{\psi}(\tau,\theta)\ ,\qquad 
M\equiv \frac{1}{4}\partial_\theta^2 + A(\theta;a/b) \partial_\theta
  +B(\theta;a/b)
\label{diff}
\end{equation}
The formal solution of this equation is given by 
$\vec{\psi}(\tau,\theta)=e^{\tau M}\vec{\psi}_\textrm{ini}(\theta)$
where $\vec{\psi}_\textrm{ini}(\theta)$
is an initial function.
For sufficiently late time, 
the solution approaches
\begin{equation}
 \vec{\psi}=e^{\tau M}\, \vec{\psi}_\textrm{ini} \to e^{\tau \alpha_0}\,
  \vec{\psi}_0\quad(\tau\to \infty)\ ,
\label{converge}
\end{equation}
where $\alpha_0$ is the largest eigen value of $M$ and $\vec{\psi}_0$ is its
eigen vector. Thus, by solving the time evolution of Eq.(\ref{diff}), we can
determine the largest eigen value of operator $M$.
The second largest eigen value can be also determined by projecting out
the $\vec{\psi}_0$ at every time step. By the similar way, we can determine
eigen values from top to bottom.

To solve the time evolution of Eq.(\ref{diff}) numerically,
we have to discretize $\tau$ and $\theta$ as $\tau_n=n\Delta \tau$ ($n=0,1,2,\cdots$) 
and $\theta_j=j\Delta\theta$ ($j=0,1,2,\cdots,N$), where 
$\Delta\theta=\pi/(2N)$.
Defining $\vec{\psi}(\tau_n,\theta_j)=\vec{\psi}^n_j$,
we discretize the Eq.(\ref{diff}) implicitly as
\begin{equation}
 \frac{\vec{\psi}^{n}_{j}-\vec{\psi}^{n-1}_{j}}{\Delta \tau}
=\frac{1}{4}\frac{\vec{\psi}^{n}_{j+1}-2\vec{\psi}^{n}_{j}+\vec{\psi}^{n}_{j-1}}{\Delta\theta^2}
+A(\theta_j)\frac{\vec{\psi}^{n}_{j+1}-\vec{\psi}^{n}_{j-1}}{2\Delta\theta}
+B(\theta_j)\vec{\psi}^{n}_{j}\ .
\label{imp}
\end{equation}
where we omitted the argument $a/b$ in matrices $A$ and $B$.
To determine $\{\vec{\psi}^{n}_{j}\}$ from $\{\vec{\psi}^{n-1}_{j}\}$, we need 
matrix inversions at every time steps.
In the implicit discretization~(\ref{imp}), 
the numerical time evolution is stable for any value of $\Delta \tau$.
Taking a large value of $\Delta\tau$ makes the convergence of time
evolution~(\ref{converge}) fast. 

At axis $\theta=0,\pi/2$, there are regular singularities.
The matrix $A(\theta)$ diverges at axis as
$A(\theta)\to A_0/\theta$ ($\theta\to 0$) and
$A(\theta)\to A_1/(\pi/2-\theta)$ ($\theta\to \pi/2$),
where $A_0$ and $A_1$ are constant matrices.
On the other hand, thanks to the definition of the variables~(\ref{redef_MXW}),
the matrix $B(\theta)$ is regular at axis.
For the regularity at axis, the variable $\vec{\psi}$ must satisfy the Neumann boundary
condition, namely,
$\vec{\psi}(\tau,\theta)=\vec{\psi}(\tau,0)+(1/2)\vec{\psi}''(\tau,0)\theta^2$
(for $\theta\to 0$) and
$\vec{\psi}(\tau,\theta)=\vec{\psi}(\tau,\pi/2)+(1/2)\vec{\psi}''(\tau,\pi/2)(\pi/2-\theta)^2$ 
(for $\theta\to \pi/2$), where $'\equiv \partial/\partial\theta$.
Thus, at axis, the equation~(\ref{diff}) reduces to
\begin{equation}
\begin{split}
&\partial_\tau\vec{\psi}(\tau,0)=\left(\frac{1}{4}+A_0\right)\vec{\psi}''(\tau,0)+B(0)\vec{\psi}(\tau,0)\ ,\\
&\partial_\tau\vec{\psi}(\tau,\pi/2)=\left(\frac{1}{4}-A_1\right)\vec{\psi}''(\tau,\pi/2)+B(\pi/2)\vec{\psi}(\tau,\pi/2)\ ,\\
\end{split}
\end{equation}
Therefore, 
we can discretize the equation at axis as
\begin{equation}
\begin{split}
&\frac{\vec{\psi}^{n}_{0}-\vec{\psi}^{n-1}_{0}}{\Delta \tau}
=\left(\frac{1}{4}+A_0\right)\frac{2\vec{\psi}^{n}_{1}-2\vec{\psi}^{n}_{0}}{\Delta\theta^2}
+B(0)\vec{\psi}^{n}_{0}\ ,\\
&\frac{\vec{\psi}^{n}_{N}-\vec{\psi}^{n-1}_{N}}{\Delta \tau}
=\left(\frac{1}{4}-A_1\right)\frac{2\vec{\psi}^{n}_{N-1}-2\vec{\psi}^{n}_{N}}{\Delta\theta^2}
+B(\pi/2)\vec{\psi}^{n}_{N}\ ,
\end{split}
\label{impb}
\end{equation}
where we used the Neumann boundary condition,
$\vec{\psi}^{n}_{-1}=\vec{\psi}^{n}_{1}$ and
$\vec{\psi}^{n}_{N+1}=\vec{\psi}^{n}_{N-1}$.
We solved the system of discretized equations~(\ref{imp}) and
(\ref{impb}) setting $\Delta \tau=1.0$ and $N=2000$.
To avoid numerical overflow, we normalized the variable as
$\sum_{j=0}^{N}|\vec{\psi}^n_j|^2=1$ at each time step.
As the result, we obtained the sequence of conformal weights as in Table.\ref{MaxMPBH}.
We checked that the result do not depend on $\Delta \tau$ and $N$ if
we chose sufficiently large $N$.


\begin{thebibliography}{99}

\bibitem{Guica:2008mu}
  M.~Guica, T.~Hartman, W.~Song and A.~Strominger,
  Phys.\ Rev.\  D {\bf 80} (2009) 124008
  [arXiv:0809.4266 [hep-th]].

\bibitem{Bredberg:2011hp}
  I.~Bredberg, C.~Keeler, V.~Lysov and A.~Strominger,
  arXiv:1103.2355 [hep-th].

\bibitem{Bardeen:1999px}
  J.~M.~Bardeen and G.~T.~Horowitz,
  Phys.\ Rev.\  D {\bf 60}, 104030 (1999)
  [arXiv:hep-th/9905099].

\bibitem{Dias:2009ex}
  O.~J.~C.~Dias, H.~S.~Reall and J.~E.~Santos,
  JHEP {\bf 0908} (2009) 101
  [arXiv:0906.2380 [hep-th]].

\bibitem{Amsel:2009ev}
  A.~J.~Amsel, G.~T.~Horowitz, D.~Marolf and M.~M.~Roberts,
  JHEP {\bf 0909} (2009) 044
  [arXiv:0906.2376 [hep-th]].

\bibitem{Maldacena:1998uz}
  J.~M.~Maldacena, J.~Michelson and A.~Strominger,
  JHEP {\bf 9902}, 011 (1999)
  [arXiv:hep-th/9812073].

\bibitem{Matsuo:2009sj}
  Y.~Matsuo, T.~Tsukioka and C.~M.~Yoo,
  Nucl.\ Phys.\  B {\bf 825}, 231 (2010)
  [arXiv:0907.0303 [hep-th]].

\bibitem{Matsuo:2009pg}
 Y.~Matsuo, T.~Tsukioka and C.~M.~Yoo,
  [arXiv:0907.4272 [hep-th]].

\bibitem{Castro:2009jf}
  A.~Castro and F.~Larsen,
  JHEP {\bf 0912}, 037 (2009)
  [arXiv:0908.1121 [hep-th]].

\bibitem{Rasmussen:2010sa}
  J.~Rasmussen,
  Int.\ J.\ Mod.\ Phys.\  A {\bf 25}, 5517 (2010)
  [arXiv:1004.4773 [hep-th]].


\bibitem{Bredberg:2009pv}
  I.~Bredberg, T.~Hartman, W.~Song and A.~Strominger,
  JHEP {\bf 1004}, 019 (2010)
  [arXiv:0907.3477 [hep-th]].

\bibitem{Cvetic:2009jn}
  M.~Cvetic and F.~Larsen,
  JHEP {\bf 0909}, 088 (2009)
  [arXiv:0908.1136 [hep-th]].

\bibitem{Hartman:2009nz}
  T.~Hartman, W.~Song, A.~Strominger,
  JHEP {\bf 1003}, 118 (2010).
  [arXiv:0908.3909 [hep-th]].


\bibitem{Castro:2010fd}
  A.~Castro, A.~Maloney, A.~Strominger,
  Phys.\ Rev.\  {\bf D82}, 024008 (2010).
  [arXiv:1004.0996 [hep-th]].

\bibitem{Kunduri:2007vf}
  H.~K.~Kunduri, J.~Lucietti and H.~S.~Reall,
  Class.\ Quant.\ Grav.\  {\bf 24} (2007) 4169
  [arXiv:0705.4214 [hep-th]].


\bibitem{Kunduri:2008rs}
  H.~K.~Kunduri and J.~Lucietti,
  J.\ Math.\ Phys.\  {\bf 50}, 082502 (2009)
  [arXiv:0806.2051 [hep-th]].

\bibitem{Hollands:2009ng}
  S.~Hollands and A.~Ishibashi,
  Annales Henri Poincare {\bf 10}, 1537 (2010)
  [arXiv:0909.3462 [gr-qc]].

\bibitem{Pomeransky:2006bd}
  A.~A.~Pomeransky and R.~A.~Sen'kov,
  arXiv:hep-th/0612005.

\bibitem{Myers:1986un}
  R.~C.~Myers and M.~J.~Perry,
  Annals Phys.\  {\bf 172}, 304 (1986).

\bibitem{Rasheed:1995zv}
  D.~Rasheed,
  Nucl.\ Phys.\  B {\bf 454}, 379 (1995)
  [arXiv:hep-th/9505038].

\bibitem{Larsen:1999pp}
  F.~Larsen,
  ``Rotating Kaluza-Klein black holes,''
  Nucl.\ Phys.\  B {\bf 575}, 211 (2000)
  [arXiv:hep-th/9909102].

\bibitem{Durkee:2010qu}
  M.~Durkee and H.~S.~Reall,
  Class.\ Quant.\ Grav.\  {\bf 28}, 035011 (2011)
  [arXiv:1009.0015 [gr-qc]].

\bibitem{Durkee:2010ea}
  M.~Durkee and H.~S.~Reall,
  arXiv:1012.4805 [hep-th].
(In this paper, there are some typos. For example, our expresions~(\ref{Oscalar}),
	(\ref{Omaxwell}), (\ref{a=b_MXW}) and (\ref{a=b_grav}) are
	different from those in~\cite{Durkee:2010ea}.
In a revised version, these typos are going to be corrected.)


\bibitem{Dias:2010eu}
  O.~J.~C.~Dias, P.~Figueras, R.~Monteiro, H.~S.~Reall and J.~E.~Santos,
  JHEP {\bf 1005}, 076 (2010)
  [arXiv:1001.4527 [hep-th]].

\bibitem{Kunduri:2006qa}
  H.~K.~Kunduri, J.~Lucietti and H.~S.~Reall,
  Phys.\ Rev.\  D {\bf 74}, 084021 (2006)
  [arXiv:hep-th/0606076].

\bibitem{Guica:2010ej}
  M.~Guica and A.~Strominger,
  JHEP {\bf 1102}, 010 (2011)
  [arXiv:1009.5039 [hep-th]].

\bibitem{Compere:2010uk}
  G.~Compere, W.~Song, A.~Virmani,
    [arXiv:1010.0685 [hep-th]].

\bibitem{Sheinblatt:1997nt}
  H.~J.~Sheinblatt,
  Phys.\ Rev.\  D {\bf 57}, 2421 (1998)
  [arXiv:hep-th/9705054].

\bibitem{Itzhaki:1998ka}
  N.~Itzhaki,
  JHEP {\bf 9809}, 018 (1998)
  [arXiv:hep-th/9809063].

\bibitem{Emparan:2006it}
  R.~Emparan and G.~T.~Horowitz,
  Phys.\ Rev.\ Lett.\  {\bf 97}, 141601 (2006)
  [arXiv:hep-th/0607023].

\bibitem{Emparan:2007en}
  R.~Emparan and A.~Maccarrone,
  Phys.\ Rev.\  D {\bf 75}, 084006 (2007)
  [arXiv:hep-th/0701150].

\bibitem{Gross:1983hb}
  D.~J.~Gross and M.~J.~Perry,
  Nucl.\ Phys.\  B {\bf 226}, 29 (1983).

\bibitem{Sorkin:1983ns}
  R.~d.~Sorkin,
  Phys.\ Rev.\ Lett.\  {\bf 51}, 87 (1983).

\bibitem{Murata:2008yx}
  K.~Murata and J.~Soda,
  Prog.\ Theor.\ Phys.\  {\bf 120}, 561 (2008)
  [arXiv:0803.1371 [hep-th]].

\bibitem{Kodama:2003jz}
  H.~Kodama and A.~Ishibashi,
  Prog.\ Theor.\ Phys.\  {\bf 110}, 701 (2003)
  [arXiv:hep-th/0305147].

\bibitem{Ishibashi:2003ap}
  A.~Ishibashi and H.~Kodama,
  Prog.\ Theor.\ Phys.\  {\bf 110}, 901 (2003)
  [arXiv:hep-th/0305185].

\bibitem{Dias:2009iu}
  O.~J.~C.~Dias, P.~Figueras, R.~Monteiro, J.~E.~Santos and R.~Emparan,
  Phys.\ Rev.\  D {\bf 80}, 111701 (2009)
  [arXiv:0907.2248 [hep-th]].


\bibitem{Morisawa:2004tc}
  Y.~Morisawa and D.~Ida,
  Phys.\ Rev.\  D {\bf 69}, 124005 (2004)
  [arXiv:gr-qc/0401100].

\bibitem{Hollands:2007aj}
  S.~Hollands and S.~Yazadjiev,
  Commun.\ Math.\ Phys.\  {\bf 283}, 749 (2008)
  [arXiv:0707.2775 [gr-qc]].

\end{thebibliography}
\end{document}